\documentclass[twocolumn,aps,prl,nopacs,superscriptaddress]{revtex4}
\usepackage{tabularx} 
\usepackage{amsmath}  
\usepackage{graphicx} 
\usepackage[margin=1in,letterpaper]{geometry} 
\usepackage[final]{hyperref} 
\hypersetup{
	colorlinks=true,       
	linkcolor=blue,        
	citecolor=blue,        
	filecolor=magenta,     
	urlcolor=blue         
}

\begin{document}

\title{Fourier-based methods for removing mesh anomalies from angle resolved photoemission spectra}
\author{Shouzheng Liu}
\author{Erica Kotta}
\author{Yishuai Xu}
\affiliation{Department of Physics, New York University, New York, New York 10003, USA}
\author{Joshua Mutch}
\author{Jiun-Haw Chu}
\affiliation{Department of Physics, University of Washington, Seattle, WA, USA}
\author{Moritz Hoesch}
\affiliation{Photon Science, Deutsches Elektronen-Synchrotron (DESY), Notkestrasse 85, 22607
Hamburg, Germany}
\author{Sanjoy Kr Mahatha}
\affiliation{School of Physics and Materials Science, Thapar Institute of Engineering and Technology, Patiala - 147004, India}
\affiliation{Photon Science, Deutsches Elektronen-Synchrotron (DESY), Notkestrasse 85, 22607
Hamburg, Germany}
\author{Jonathan D. Denlinger}
\affiliation{Advanced Light Source, Lawrence Berkeley National Laboratory, Berkeley, CA 94720, USA}

\author{L. Andrew Wray}
\email{lawray@nyu.edu}
\thanks{Corresponding author}
\affiliation{Department of Physics, New York University, New York, New York 10003, USA}

\begin{abstract}

Recent improvements to spatial resolution in angle-resolved photo-emission spectroscopy (ARPES) have made it common to perform measurements with a very brief dwell time, for the purpose of mapping the spectral function over large surface regions. However, rapid measurement modalities can suffer a grid-like intensity modulation due to a wire mesh that is typically placed in front of the ARPES detector to block stray electrons. Here, we explore Fourier-based methods that can effectively remove this artifact, and improve the quality of ARPES images obtained in rapid scanning modes. An open source software package is provided containing implementations of demonstrated algorithms. 
\end{abstract}

\maketitle

\section{Introduction}
\begin{figure}[ht]
    \centering
    \includegraphics[width=8cm]{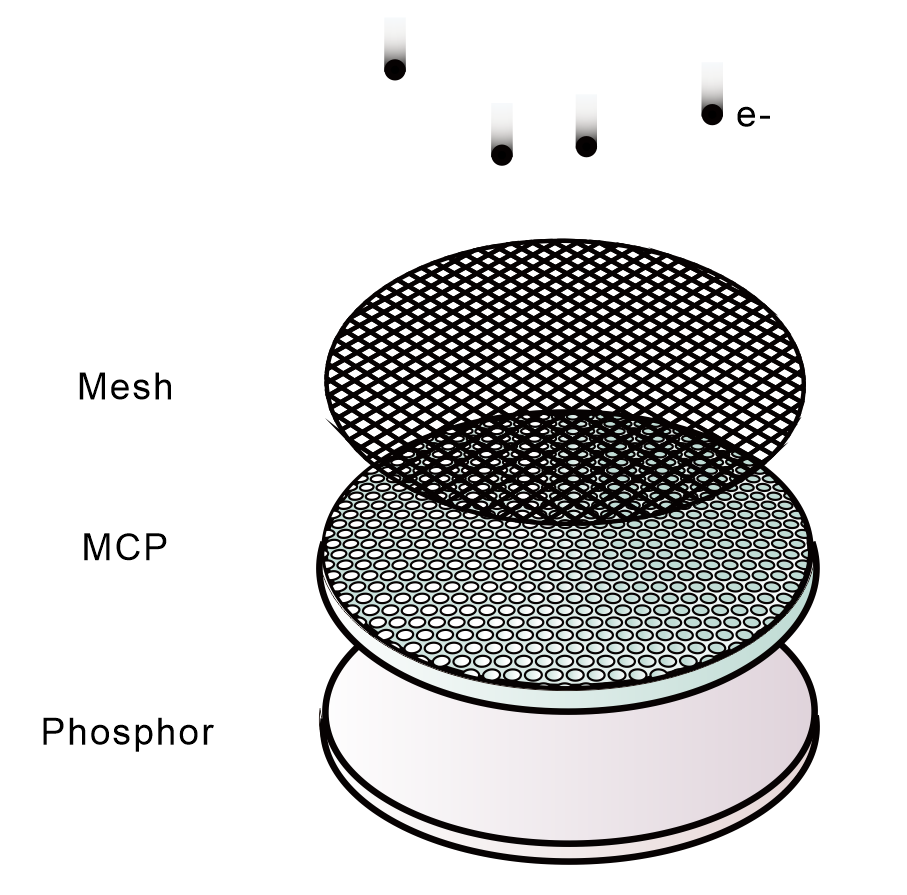}
    \caption{Electrons incident on an ARPES detector.  Image shows the stacking order of the phosphor, MCP, and gold mesh layers.}
    \label{fig:detector}
\end{figure}

The recent worldwide development of spectromicroscopy beamlines for angle resolved photoemission (ARPES) has introduced a new measurement regime in which it is often necessary to take extremely rapid measurements. While traditional high-quality ARPES images are often recorded on the scale of 10s of minutes, spectromicroscopy measurements are often limited to the $\sim$1 second time scale \cite{ericaPaper,UltimateResolution} due to the need to map large surface regions, and to avoid sample damage from a highly focused beam. One challenge when performing measurements on such a rapid time scale is that a gold mesh often incorporated in ARPES detectors (see Fig. \ref{fig:detector}) can cause grid-like patterns to appear on scan images that captured with the most rapid data acquisition modes (``Fixed" or ``Dithered" modes). These mesh patterns can vary significantly as a function of position in large spatial scans, meaning that corrections must be generated dynamically and are measurement-specific. Here, we explore the application of digital image processing methods to correct this type of artifact.
\par
A very similar problem has been previously addressed in the removal of Moir\'e patterns that appear in photographs of computer screens \cite{moireRemove1}\cite{moireRemove2}. As with the grid-like patterns appear on ARPES scan images, these these Moir\'e patterns constitute an undesired periodic intensity modulation, and the same classes of removal procedures can generally be applied to both scenarios. Nevertheless, ARPES images constitute a very specific use case.
\par
In this work, we will first demonstrate that the intensity modulation of broader features in ARPES images can be easily removed with a Fourier-based method. A mesh template can be created in this process and be used on other scan images in the same beamline. We will also explore a more computationally challenging method that can be used to minimize the appearance of measurement artifacts on sharper spectral features, with widths similar to the mesh periodicity. Related Python code can be accessed via the link in Ref. \cite{githubLink}.
\subsection{Structure of the ARPES detector}
In a commercial ARPES system, the detector is usually composed of 4 parts. The first layer is a wire mesh, following by a multichannel plate (MCP), a layer of phosphor, and a camera. A bias voltage is applied to the mesh to block background electrons. The MCP plate has a honeycomb structure, in which each channel serves as an amplifier. When electrons collide with the wall of the channel, they produce a large number of secondary electrons. The phosphor, excited by secondary electrons, will then produce detectable light signals for the camera. 
\par
The wire mesh casts a shadow that modulates the intensity of the scanned image. Intensity of the shadow can be mitigated by fine-tuning the electron optics (particularly the Scienta Detector Retard (DRET) voltage) for each pass energy, and possibly as a function of angular deflection within a DA30. Unlike Moir\'e patterns discussed above that usually lack a strict periodicity, ARPES mesh patterns are highly regular, reflecting the lattice structure of the wire mesh. This feature enables us to start with a Fourier based method and achieve superior results for ARPES scan image mesh patterns removal.

Soft X-ray angle-resolved photoemission spectroscopy (SX-ARPES) measurements presented in this paper were performed at the ASPHERE III endstation at the Variable Polarization XUV Beamline P04 of the PETRA III storage ring at DESY (Hamburg, Germany). The primary data image examined in this paper was obtained with a photon energy $h\nu=260$ eV and a pass energy of 100 eV, and was selected due to containing a combination of narrow and broad ARPES features. Fig. 4-5 include averaged images from an incident energy dependence series, with incident photon energies evenly distributed from 260-360 eV. Single crystals of ZrTe$_5$ were cleaved \emph{in situ}, and data were were collected using a Scienta DA30-L analyzer at a sample temperature of $\sim$200 K and in ultra-high vacuum $<3\times10^{-10}$ mbar. The angle and energy resolution of the ARPES measurements were $\Delta\theta\sim0.1^\circ$ and $\Delta E\sim50$ meV.

\subsection{Discrete FT method}

Procedures for moir\'e-like feature removal are typically carried out in Fourier space \cite{moireRemove1}. A discrete Fourier transform (FT) of a 2D image can be performed with the following equation:
\begin{equation}
    F(k,l)=\sum_{i=0}^{N-1} \sum_{j=0}^{N-1} f(i,j)e^{-i2\pi (\frac{k i}{N}+\frac{l j}{N})}
\end{equation}
in which $f(i,j)$ is the original 2D image, which for simplicity is taken to be square and $N\times N$ pixels in size. The discrete FT discomposes the original spatial image into a Fourier basis with amplitude and phase factors described by the coefficient $F(k,l)$. Typical properties of ARPES discrete FT images are discussed in Ref. \cite{LiFT_SDI}.

Removing Fourier peaks associated with the mesh periodicity has the same effect as removing corresponding period structures in the original image. After applying filters to frequency domain FT images, the corrected spatial domain image is obtained through an inverse FT. Procedural details specific to ARPES are outlined in the subsections below.

\section{Removing mesh peaks in Fourier space}

\subsection{Fixing mesh distortion}
Though the mesh pattern in ARPES images is expected have a structure that closely resembles the wire mesh, data acquisition software corrections that flatten the Fermi level can leave the mesh warped within data images. The distorted mesh pattern results in broad and abnormally shaped peaks, which can be seen along the diagonal axes of Fig. \ref{fig:curvature}(b). Removing these broad peaks has a risk of removing other useful features from images. Hence it is helpful to remap the image with linear shifts along the energy axis. For Fig. 2(c-d) and all successive analyses in the text, this has been accomplished by selecting a radius of curvature $R_0$ and distorting the energy axis to match that curvature as follows:

\begin{figure}[ht]
    \centering
    \includegraphics[width=8.1cm]{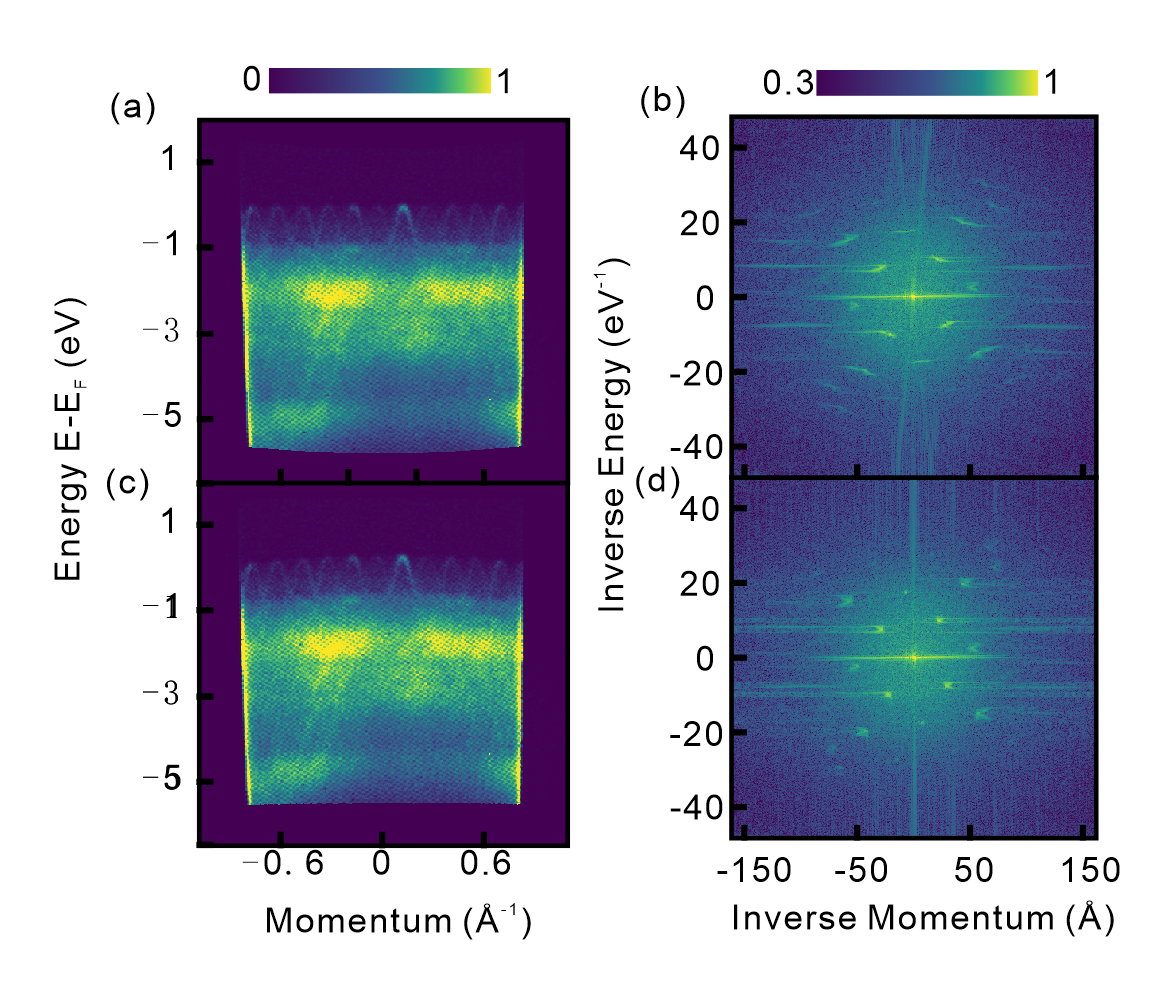}
    \caption{Obtaining a sharp mesh image in Fourier space. (a) Original ARPES detector image (cropped). (b) The Fourier transform of panel (a). (c) A modified version of the ARPES image, in which Eq. (2) has been used to remove the distortion of the mesh pattern. (d) Fourier transform of the modified panel (c) data, showing sharper mesh-associated Fourier peaks near the diagonals of the image.}
    \label{fig:curvature}
\end{figure}


\begin{gather}
    f'\left(i,j\right)=f\left(i,j+R_0 (cos(\theta(i))-cos(\theta(i_c))\right) \\
    \theta(i)=asin((i_c-i)/R_0)
\end{gather}
where $i_c$ indicates the horizontal center of the image. Non-integer indices are interpreted via interpolation. The radius parameter $R_0$ is selected to maximize amplitude (and thus sharpness) of the mesh-associated peaks in the Fourier transform. The distorted image ($f'(i,j)$) is shown in Fig. \ref{fig:curvature}(c), and has visibly sharper mesh-associated features within a Fourier transform (Fig. \ref{fig:curvature}(d)). The energy-axis shifts must be reversed after mesh-feature removal to restore a flat Fermi level.

\subsection{Removing mesh Fourier peaks}
\begin{figure*}[ht]
    \centering
    \includegraphics[width=14cm]{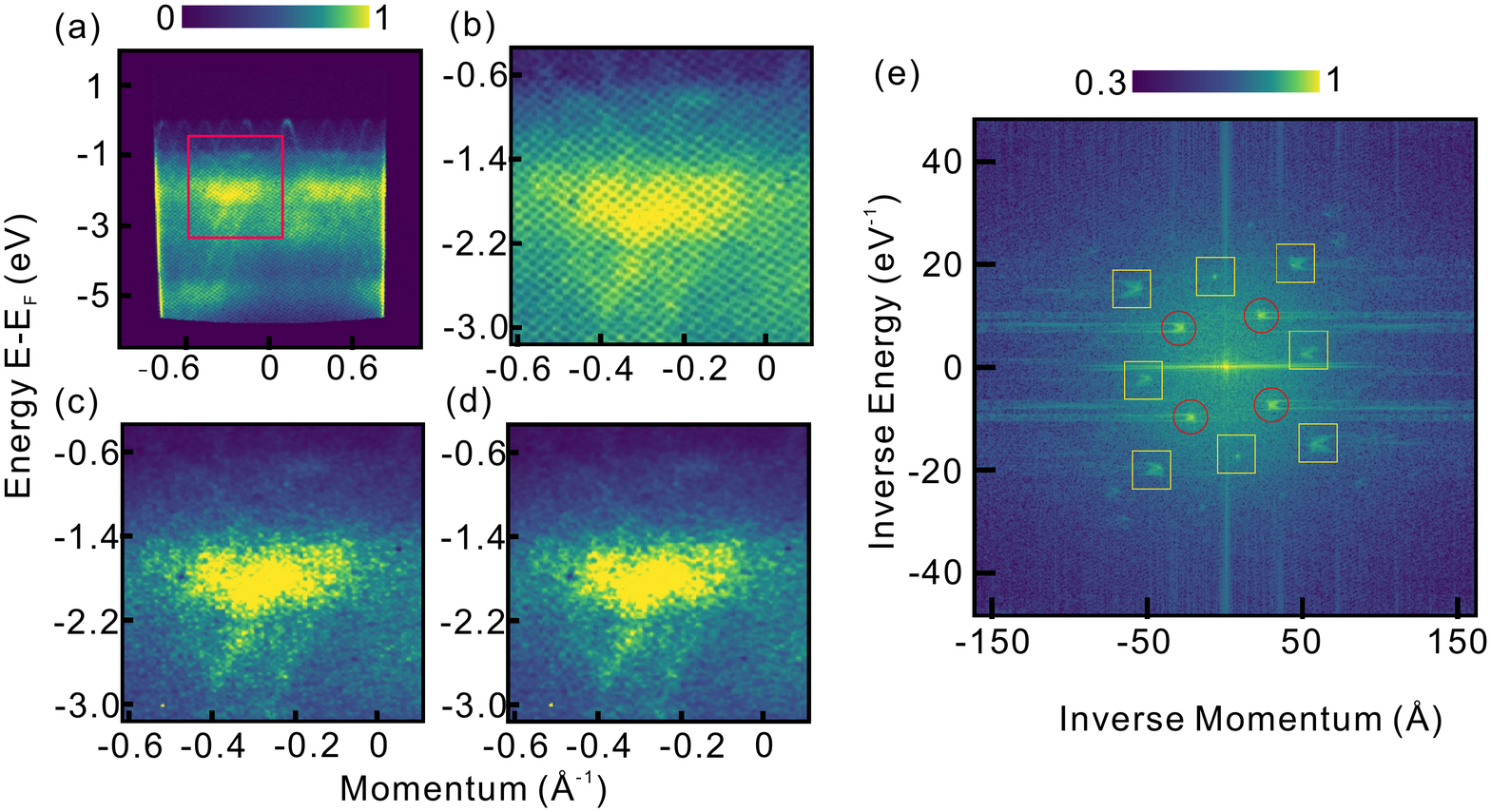}
    \caption{Removing first and second order mesh features.  (a) Original scan image from ARPES. (b) Close-up showing the mesh pattern in the red box region indicated in panel (a). (c) Close-up after removing the first order mesh Fourier peaks. (d) Close-up after removing both first and second order peaks. (e) Fourier transform of panel (a), with warping corrected by Eq. (2). First order peaks indicated with red circles and second order peaks with yellow boxes.}
    \label{fig:removepeaks}
\end{figure*}

The mesh pattern can be clearly seen by eye a close-up of the ARPES data (Fig. \ref{fig:removepeaks}(b)). Four first order Fourier peaks representing reciprocal vectors of the mesh are indicated with circles in Fig. \ref{fig:removepeaks}(e), as are eight second order peaks obtained as the sum of two reciprocal vectors. 

For this manuscript, mesh peaks were removed by surveying the surrounding amplitudes to define a maximum amplitude threshold, and setting the `mesh' pixels that exceeded this intensity to equal the mean of the local region. However, this procedure was not refined, as for this particular data set no dramatic visual advantage was identified with different tested approaches that included: downscaling `mesh' pixel amplitude to match the background while preserving phase; setting the intensity of `mesh' pixels to zero; removing an inverted Gaussian centered on the mesh peak.

\par
Panels (c-d) of Fig. \ref{fig:removepeaks} separately show the result of removing just the first order peaks, and removing both first and second order peaks for relatively featureless portion of the ARPES image. A close examination of the two panels reveals that removing the first order peaks leaves a grainy background in which the grid periodicity can still be observed. The grid periodicity is not readily recognizable when both first and second order Fourier peaks have been removed.

The success of this approach is unsurprising so long as the ARPES image contains no real features with a width that resembles the mesh periodicity, so that the mesh features can be cleanly isolated from real spectral features within the Fourier transform. When this is not the case, one should consider obtaining a mesh template from a more uniform image (see below) or applying a more sophisticated feature removal algorithm (see next section).

\subsection{Creating a mesh template}
When mesh corrections will be applied to a wide range of measurements rather than just a single image, it is useful to generate a template matrix that encodes the mesh pattern and can apply the correction to multiple ARPES images measured under similar conditions. In the data set addressed in this paper, we find that a common mesh template is effective for an incident photon energy dependence series in which all other measurement parameters are held constant. The mesh template $T(i,j)$ is defined by
\begin{equation}
    T(i,j)=\frac{I(i,j)}{I_0(i,j)}.
\end{equation}
Here, $I_0(i,j)$ is the original scan image with intensity modulation due to the wire mesh, while $I(i,j)$ is a corrected image obtained by removing the Fourier peaks of the mesh as described earlier. Mesh corrections can be applied to a new image $I_1$ by taking the Hademard product $T(i,j)I_1(i,j)$.
\par
\begin{figure*}[ht]
    \centering
    \includegraphics[width=14cm]{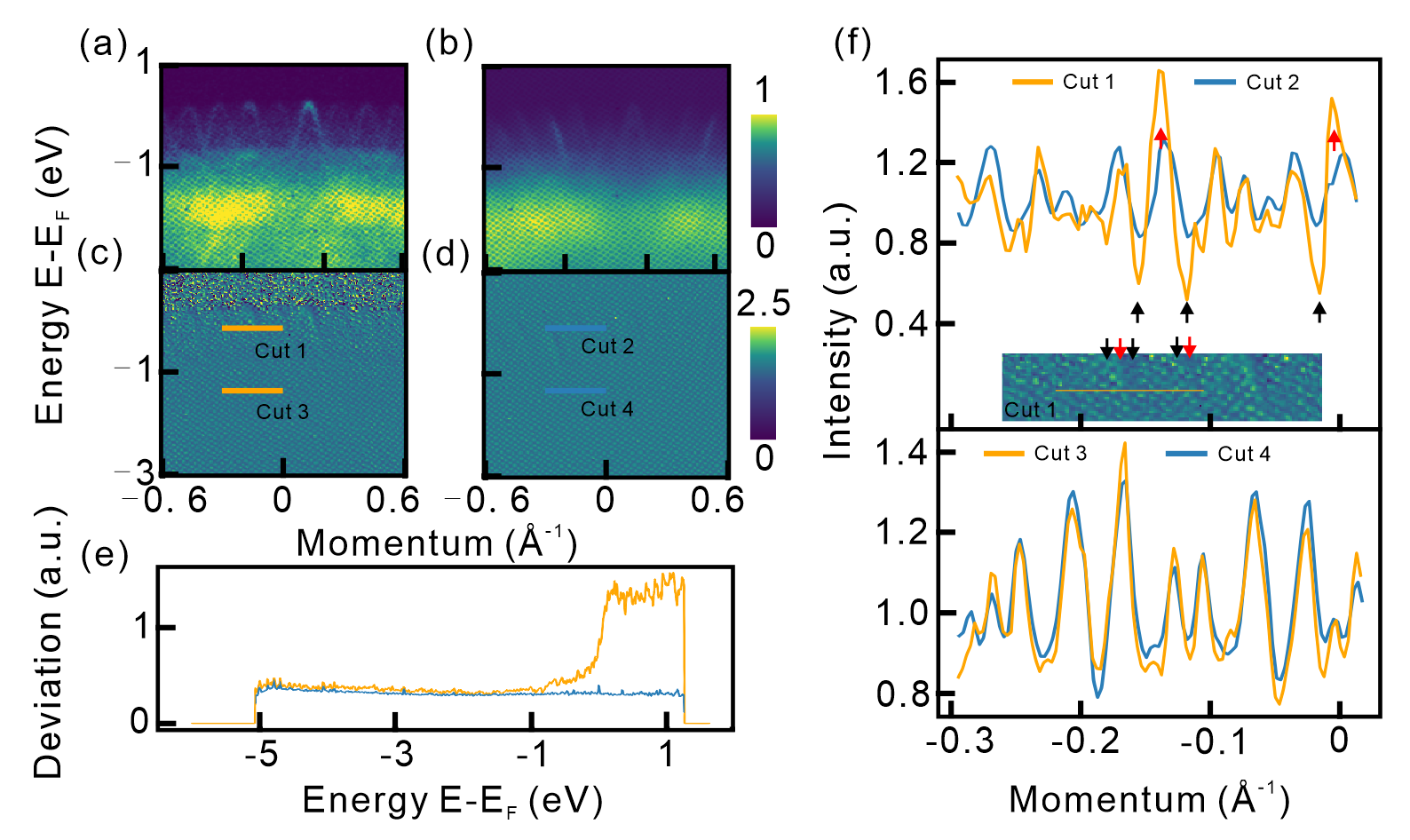}
    \caption{Templates for mesh correction. (a) A cropped section of a single ARPES image. (b) The average of 51 images obtained at different incident photon energies, with the same pass energy and measurement geometry. (c-d) Mesh correction templates ($M_0$) for panels (a-b), generated via the discrete FT method. (e) The standard deviation of constant-energy contours in (yellow) the single image template and (blue) the averaged image template within the cropped region. (f) Constant-energy mesh template contours are shown for the same momentum region from the single image template (cuts 1 and 3) and the averaged image template (cuts 2 and 4).}
    \label{fig:avgmesh}
\end{figure*}

An example of an ARPES image and the mesh template obtained from it is shown in Fig. \ref{fig:avgmesh}(a,c), and is useful to examine in detail. As noted in the previous subsection, the mesh template suffers imperfections where it overlaps with sharp band features that approach the mesh periodicity. This issue is particularly severe when the bands disperse along a trajectory that is orthogonal to a reciprocal lattice vector of the mesh, meaning that a large component of the Fourier transform of the band will overlap with the reciprocal space Fourier peak of the mesh. Moreover, limited measurement statistics above the Fermi level render the mesh template useless in that region.

These failures of the template can be largely corrected by using a larger data set. Figure \ref{fig:avgmesh}(b) presents an averaged ARPES image obtained from 51 images measured at photon energies evenly distributed from $h\nu=260-360$ eV. Averaging over the z-axis dispersion of the sample results in broader features that leave less pronounced scarring on the template (see Fig. \ref{fig:avgmesh}(d)). Moreover, this larger data set provides adequate sampling above the Fermi level.

The improvement can also be seen by comparing the standard deviation of the two mesh templates as a function of energy (Fig. \ref{fig:avgmesh}(e)). The standard deviation represents the amplitude of the mesh pattern and of fluctuations from error and noise, which add as orthogonal vectors. The averaged spectrum yields a standard deviation curve (blue curve) that is visibly far more regular, and converges with the highest quality region of the yellow single-image curve (found at roughly 1.5-2 eV binding energy).

Figure \ref{fig:avgmesh}(f) provides a closer comparison between constant-energy contours of the two templates. When examining near the Fermi level (cuts 1 and 2), the single-image template shows dramatic deviations accounting for up to $\sim 30\%$ of spectral intensity, which correspond to blemishes associated with sharp band features. Deeper binding energies lack sharp features within the data set, as is expected from basic theoretical considerations, and show typical deviations of $<5\%$ (see cuts 3 and 4). The improved correction provided by the averaged-spectrum template is also demonstrated in Fig. \ref{fig:fig5}, discussed below.

\section{Continuous Fourier space method to improve mesh correction for sharp spectral features}

The Fourier-based methods presented above constitute an efficient way to remove mesh patterns from Fixed Mode ARPES scan images. However, the manipulation of Fourier space can result in the distortion of sharper band features, when their width and angular alignment correspond with the mesh periodicity. This section will briefly discuss an approach for using the accurate mesh template obtained from homogeneous spectral regions to generate more accurate corrections in regions with sharp features.

\begin{figure*}[ht]
    \centering
    \includegraphics[width= 16cm]{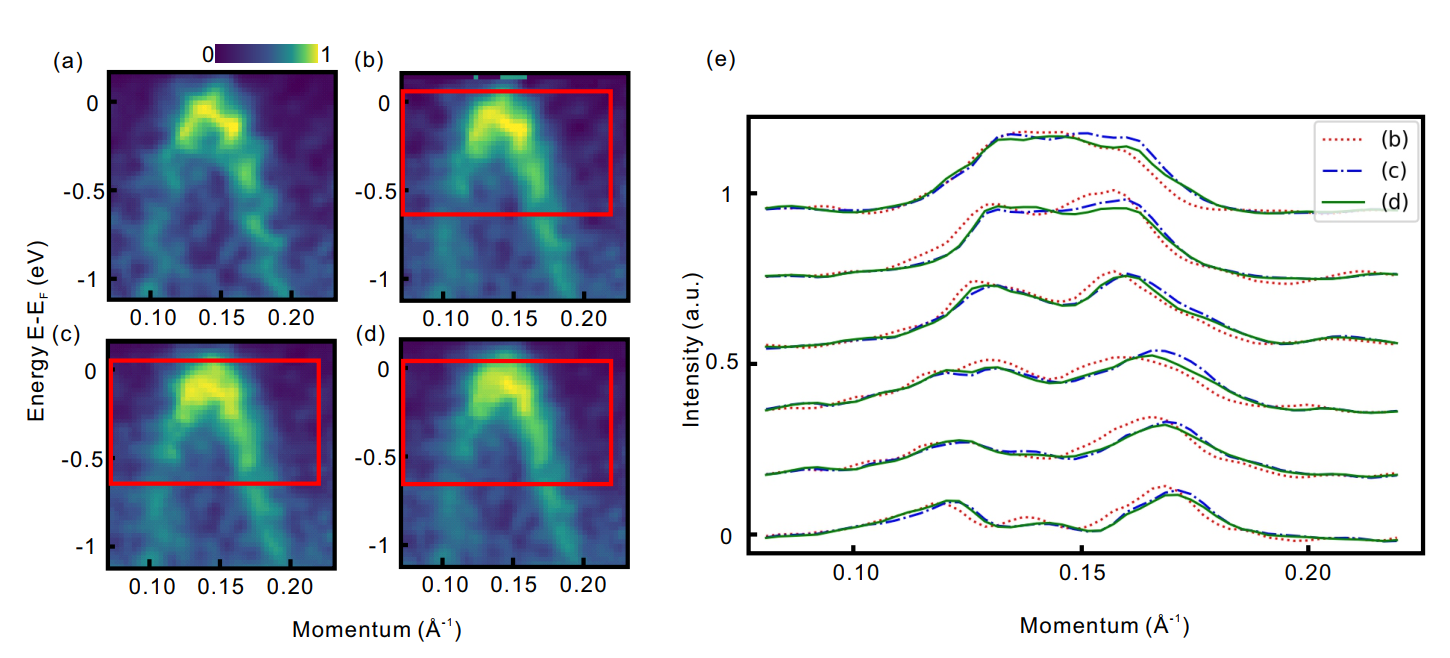}
    \caption{(a) An uncorrected ARPES image of the sharply resolved ZrTe$_5$ valence band. (b) A corrected ARPES image obtained from the discrete FT method on the single scan image ($M_0$). (c) A corrected image obtained from the discrete FT method, using the average of an incident energy dependence series to obtain the mesh template. (d) A corrected image obtained from the continuous Fourier space method. (e) Momentum distribution curves covering the red box in panels (b-d).}
    \label{fig:fig5}
\end{figure*}

We begin by applying the methods from the previous section to obtain an approximate template for the mesh ($M_0$). An algorithm for improving on $M_0$ is based on the following series of decisions: \\
1. As the mesh is globally non-uniform, the template is broken into a large number of overlapping rectangular regions indexed by $i$ ($M_{0}^i$) that can be separately addressed, and in which we expect the mesh to be locally periodic. \\
2. Because each individual subregion $M_{0}^i$ is small, a discrete FT suffers severe boundary artifacts, and cannot capture the precise angle of the mesh. To mitigate this, we switch to a continuous Fourier representation to describe the improved mesh template ($M_1$) within each subregion ($M_{1}^i$), with nonzero components only found at reciprocal lattice vectors of the mesh. \\
3. An optimization metric and procedure are introduced to obtain the continuous Fourier space representation of the template in each subregion. Constraints may be imposed to improve this result, such as forcing continuity or monotonic evolution of parameters. \\
4. Mesh templates of the sub-regions are recombined to obtain a template for the full ARPES image. The overlapping boundary regions are averaged. In our implementation, we have applied a weighted average based on the quality of the step 3 optimization in each region. For Fig. \ref{fig:fig5}(d), the algorithm was applied with square $100\times100$ pixel subregions, with horizontal and vertical step sizes of 50 pixels. As such, each pixel in the mesh template used for Fig. \ref{fig:fig5}(d) comes from the average of 4 subregions.


The result of this procedure is presented in Fig. \ref{fig:fig5}. Clear mesh-pattern artifacts can be observed on top of a sharp band feature in the $M_0$ corrected image in Fig. \ref{fig:fig5}(b), and are largely eliminated in the continuous Fourier space correction method described in this section (Fig. \ref{fig:fig5}(d)). This result is quantitatively similar to a corrected image obtained from an exhaustively hv-averaged mesh template (see Fig. \ref{fig:fig5}(c), and curve comparison in Fig. \ref{fig:fig5}(e)). This method can also provide a mesh template for spectral regions with limited statistics, such as points above the Fermi level. Steps 2-3 of the algorithm are nontrivial, and will be elaborated on below.

\subsection{Optimization of the continuous Fourier space method}

As with discrete FT procedures in the previous section, the continuous Fourier representation considered in this section considers only first- and second-order reciprocal vectors of the mesh. These vectors are defined in terms of the primitive reciprocal lattice vectors $\vec{b_1}$ and $\vec{b_2}$ as $\vec{G}_{m,n}=m\vec{b_1}+n\vec{b_2}$, with interger indices $0<|m|+|n|\leq 2$.

The amplitude and relative phases of these indices are initialized based on the amplitudes and phases associated with each reciprocal lattice vector in the full-image Fourier transform used to obtain $M_0$. They are then optimized via the simplex algorithm, minimizing a normalized cross-correlation loss function:

\begin{gather}
    L_1({I'_0}^i,{I'}^i)=1- \sum_{j,k}{I'_0}^i(j,k){I'}^i(j,k)
\end{gather}

This loss function is the normalized cross correlation between ARPES images for each subregion, corrected by the original mesh ($M_0^i$) and by the improved mesh ($M_1^i$). As defined earlier, these mesh-corrected images are created via a Hademard product with the mesh template, 

\begin{gather}
    I'^i=norm({I_0}^i\odot M^i),\\
    {I'_0}^i=norm({I_0}^i\odot {M_0}^i),
\end{gather}

where $I_0^i$ is the uncorrected ARPES image within subregion $i$. The normalization function for a sub-region indexed by $i$ is a standard vector normalization ($norm(X^i)=\frac{X^i}{\sqrt{\sum_{j,k}X^i(j,k)^2}}$), rendering the loss function equivalent to optimization of an inner product between two quantum mechanical state vectors. Finally, the sum of all elements in $M^i$ is set equal to the sum of all elements in ${M_0}^i$, to eliminate the total amplitude degree of freedom retained due to use of the $norm()$ function.

The reason for basing the loss function on corrected images ($L_1({I'_0}^i,{I'}^i)$) instead of directly comparing the mesh templates ($L_1({M_0}^i,{M}^i)$) is to disregard regions with inadequate statistics in the $M_0$ template, such as above the Fermi level. In defining the $M$ template in terms of Fourier space delta functions, one makes it impossible for localized anomalies to occur in the neighborhood of sharp band features. Based on these properties, we expect the new template $M$ to be superior to the original template $M_0$ both in noisy regions and in regions for which sharp features create localized artifacts in $M_0$.

In some cases, we have found that $L_1$-optimized mesh templates can leave some residual mesh intensity in relatively flat spectral regions.  Where necessary, this can be corrected via a second optimization step using the following loss function:

\begin{gather}
    \Tilde{I}=DFT(I'^i),\\
    L_2(I'^i)=\sum_{\{\vec{G}\}}|\Tilde{I}(\vec{G})|^2.
\end{gather}

Here, a discrete FT $DFT()$ is applied to the corrected ARPES image $I'^i$, and remnant mesh intensity is identified by evaluating amplitude at the mesh vectors. As the mesh Fourier vectors $G_{m,n}$ are incommensurate with the discrete FT coordinates, the value of $\Tilde{I}(\vec{G})$ must be obtained from interpolation and/or by summing over a small local region. In our tests, we have held the $G_{m,n}$ coordinates constant during $L_2$ optimization and not attempted simultaneous optimization of a combined metric (such as $L_1+L_2$), as the $L_2$ term can be incorrectly minimized by mis-aligning the mesh Fourier vectors.

\section{Discussion and Summary}

The removal of mesh artifacts from fixed-mode and dither-mode ARPES images is beneficial for ARPES spectromicroscopy, and generally needs to be fine tuned whenever a new beam spot or pass energy is investigated. We have explored the application of Fourier based approaches for achieving this.  A straightforward procedure involving the removal of mesh features from a discrete FT is shown to be effective for smooth spectral regions with good measurement statistics. Useful operations that are specific to the ARPES instrumentation are discussed, such as the correction of warping in the mesh pattern, and the use of averaging over multiple incident photon energies to eliminate artifacts from sharp features. We further demonstrate the viability of an improved method applied in continuous Fourier space, which can greatly improve the correction for sharper features and for spectral regions with poor statistics.  

\textbf{Acknowledgements:} L.A.W. acknowledges the support of the National Science Foundation under grant No. DMR-2105081. We acknowledge DESY (Hamburg, Germany), a member of the Helmholtz Association HGF, for the provision of experimental facilities. Parts of this research were carried out at PETRA III, and we would like to thank the Kiel and W\"urzburg teams for assistance in using the ASPHERE III endstation at beamline P04. Funding for the photoemission spectroscopy endstation (Contracts 05KS7FK2, 05K10FK1, 05K12FK1, and 05K13FK1 with Kiel University; 05KS7WW1 and 05K10WW2 with W\"urzburg University) by the Federal Ministry of Education and Research (BMBF) is gratefully acknowledged. Material synthesis at the University of Washington is supported by the Gordon and Betty Moore Foundation EPiQS Initiative, grant GBMF6759 to J.-H.C, and as part of Programmable Quantum Materials, an Energy Frontier Research Center funded by the US Department of Energy (DOE), Office of Science, Basic Energy Sciences (BES), under award DE-SC0019443. This research used resources of the Advanced Light Source, a U.S. DOE Office of Science User Facility under Contract No. DE-AC02-05CH11231. 


\end{document}